\documentclass[aps,prl,superscriptaddress,twocolumn]{revtex4-1}

\usepackage{graphics}
\usepackage{graphicx} 
\usepackage{dcolumn}
\usepackage{bm}
\usepackage{epstopdf}

\begin{document}


\title{Spin-orbit driven magnetic insulating state with ${J}_{\mathrm{eff}}\mathbf{=}1/2$ character in a 4d oxide}



\author{S.~Calder}
\email{caldersa@ornl.gov}
\affiliation{Quantum Condensed Matter Division, Oak Ridge National Laboratory, Oak Ridge, Tennessee 37831, USA}

\author{L.~Li}
\affiliation{Department of Materials Science and Engineering, University of Tennessee, Knoxville, Tennessee 37996, USA}

\author{S.~Okamoto}
\affiliation{Materials Science and Technology Division, Oak Ridge National Laboratory, Oak Ridge, Tennessee 37831, USA}

\author{Y.~Choi}
\affiliation{Advanced Photon Source, Argonne National Laboratory, Argonne, Illinois 60439, USA}

\author{R.~Mukherjee}
\affiliation{Department of Materials Science and Engineering, University of Tennessee, Knoxville, Tennessee 37996, USA}

\author{D.~Haskel}
\affiliation{Advanced Photon Source, Argonne National Laboratory, Argonne, Illinois 60439, USA}

\author{D.~Mandrus}
\affiliation{Department of Materials Science and Engineering, University of Tennessee, Knoxville, Tennessee 37996, USA}
\affiliation{Materials Science and Technology Division, Oak Ridge National Laboratory, Oak Ridge, Tennessee 37831, USA}

 

\begin{abstract}
The unusual magnetic  and electronic ground states of 5d iridates has been shown to be driven by intrinsically enhanced spin-orbit coupling (SOC). The influence of appreciable but reduced SOC in creating the manifested magnetic insulating states in 4d oxides is less clear, with one hurdle being the existence of such compounds. Here we present experimental and theoretical results on Sr$_4$RhO$_6$ that reveal SOC dominated behavior. Neutron measurements show the octahedra are both spatially separated and locally ideal, making the electronic ground state susceptible to alterations by SOC. Magnetic ordering is observed with a similar structure to an analogous  ${J}_{\mathrm{eff}}\mathbf{=}1/2$ Mott iridate. We consider the underlying role of SOC in this rhodate with density functional theory and x-ray absorption spectroscopy and find a magnetic insulating ground state with ${J}_{\mathrm{eff}}\mathbf{=}1/2$ character.
 \end{abstract}
 
\pacs{75.70.Tj, 78.70.Nx, 71.20.Ps, 71.20.Be}

\maketitle

The role of relativistic spin-orbit coupling (SOC) in creating the diverse behavior of transition metal oxides (TMOs) has largely been considered a perturbation. While this has proven a valid approximation when describing 3d based TMOs the recent intense interest in 5d systems has been driven specifically by the increased effects of SOC \cite{annurev-conmatphys-020911-125138}. In 5d TMOs SOC competes with increased orbital overlap (altered bandwidth), reduced on-site Coulomb interactions ($U$) and enhanced crystal field splitting. The resulting behavior includes potential realizations of Weyl semi-metals \cite{NaturePesin, PhysRevB.83.205101}, Kitaev physics \cite{PhysRevLett.102.017205}, topological insulators \cite{topoliridate} and routes to unconventional superconductivity \cite{PhysRevLett.108.177003, Kim11072014}. 

One of the most dramatic and well studied manifestations of SOC in 5d systems was first observed in Sr$_2$IrO$_4$ with the creation of a magnetic ${J}_{\mathrm{eff}}\mathbf{=}1/2$ Mott insulating state \cite{PhysRevLett.101.076402, KimScience}. This ground state emerges in the Ir$^{4+}$ ion with 5d$^5$ occupancy due to the t$_{2g}$ manifold, in the limit of cubic crystal field splitting, being split by SOC into a filled ${J}_{\mathrm{eff}}\mathbf{=}3/2$ band and a half filled ${J}_{\mathrm{eff}}\mathbf{=}1/2$ band, with even the reduced $U$ in 5d systems being able to drive the opening of the insulating band gap via the Mott mechanism. Subsequent investigations on iridates have led to a growing list of candidate ${J}_{\mathrm{eff}}\mathbf{=}1/2$ magnetic materials  \cite{annurev-conmatphys-020911-125138}. Observations of ${J}_{\mathrm{eff}}\mathbf{=}1/2$ magnetic insulating behavior outside of the iridates is limited, with no known examples in 4d oxide based systems. Of particular interest in SOC enhanced magnetic states is the bond directional anisotropy of the ${J}_{\mathrm{eff}}\mathbf{=}1/2$ pseudo-spins and how alterations from the ideal case, such as tetragonal distortions or pressure-driven increase in bandwidth in Sr$_2$IrO$_4$ that causes mixing with ${J}_{\mathrm{eff}}\mathbf{=}3/2$ bands \cite{PhysRevLett.109.027204}, influences the creation of exotic ground states. 

The limit of a pure ${J}_{\mathrm{eff}}\mathbf{=}1/2$ state requires strong SOC and a cubic CEF environment. While 5d iridates fall into the strong SOC regime no current examples have a cubic environment. Nevertheless even with appreciable distortions the state is still realized in 5d systems, albeit with potential mixing in of the ${J}_{\mathrm{eff}}\mathbf{=}3/2$ bands \cite{annurev-conmatphys-020911-125138}.  The apparent robustness of the ground state indicates that the converse would be applicable;  ${J}_{\mathrm{eff}}\mathbf{=}1/2$ behavior should be manifested in systems with near ideal octahedra even if the SOC is reduced from those found in 5d systems. However in 4d TMOs where single-ion SOC is of the order 0.15 eV \cite{PhysRevB.86.125105, doi:10.1021/ic402653f}, compared to 0.5 eV for iridates, it is unclear whether ${J}_{\mathrm{eff}}\mathbf{=}1/2$-like Mott magnetism is manifested or whether they fall into the class of three Kramers states \cite{doi:10.1021/ic402653f}. Outside oxides, examples of SOC enhanced behavior in 4d systems are limited to theoretical predictions in fluoride based paramagnetic Ir and Rh systems, characteristic behavior in the chloride $\alpha$-RuCl$_3$ and non-magnetic semiconducting behavior in Li$_{2}$RhO$_{3}$ \cite{PhysRevLett.114.096403,PhysRevB.90.041112,ArnabRuCl3,PhysRevB.87.161121}.

Here we show a 4d-based oxide compound to exhibit a magnetically ordered ground state with ${J}_{\mathrm{eff}}\mathbf{=}1/2$ character that offers a unique viewpoint on SOC enhanced behavior in general. The material we focus on, Sr$_4$RhO$_6$, hosts a Rh valence of 4+ (4d$^5$), the same valence as observed in the ${J}_{\mathrm{eff}}\mathbf{=}1/2$ iridates (Ir$^{4+}$, 5d$^5$). Previous investigations of Sr$_4$RhO$_6$ have been extremely limited \cite{Randall:a02576,VenteSr416}. Vente {\it et al.} observed an anomaly in the susceptibility around 8 K that suggested Sr$_4$RhO$_6$ was the first magnetically ordered Rh$^{4+}$ compound, however no microscopic long range magnetic order has been measured. 


Sr$_4$RhO$_6$ forms the same hexagonal space group $R\overline{3}c$ adopted by the ${J}_{\mathrm{eff}}\mathbf{=}1/2$ iridate Ca$_4$IrO$_6$. Although not having an $O_h$ point group Ca$_4$IrO$_6$ was shown to reside close to the local cubic limit required for an unmixed ${J}_{\mathrm{eff}}\mathbf{=}1/2$ state \cite{PhysRevB.89.081104}. We directly compare the properties and the role of SOC in these related 4d and 5d materials. Structurally we find the rhodate to reside even closer to the local cubic limit and, since it is isostructural to Ca$_4$IrO$_6$, has disconnected RhO$_6$ octahedra. This makes Sr$_4$RhO$_6$ an  appealing candidate to look for ${J}_{\mathrm{eff}}\mathbf{=}1/2$ behavior since the expected narrow bands and strong 10$Dq$ splitting will help to overcome the reduced SOC in going from 5d to 4d systems. We consider these factors in contrast to Sr$_2$RhO$_4$, the rhodate analogue of Sr$_2$IrO$_4$, where the role of SOC is considered to be negligible \cite{PhysRevB.86.125105}. Sr$_2$RhO$_4$ is a paramagnetic metal where the large bandwidth and distorted octahedra, as similarly found in Sr$_2$IrO$_4$, suppress the influence of SOC \cite{PhysRevB.86.125105}. For Sr$_4$RhO$_6$ we present experimental neutron and x-ray results along with  detailed density functional theory calculations that show SOC plays a crucial role in creating a ground state with ${J}_{\mathrm{eff}}\mathbf{=}1/2$ character.


Polycrystalline samples of Sr$_4$RhO$_6$ were grown following a similar method described in the literature \cite{VenteSr416}. A stoichiometric mixture of high-purity SrCO$_3$, Rh$_2$O$_3$ were ground, pressed into pellets, and sintered in pure oxygen atmosphere at 900$^\circ$C, 1000$^\circ$C, and 1100$^\circ$C with intermediate grindings. The reaction time for each temperature was 4 days. The sample was finally annealed at 1250$^\circ$C for 15 days. The extensive heating is required to fully remove the Sr$_2$RhO$_4$ phase. A pellet was measured with a PPMS systems to probe the resistivity. Neutron powder diffraction (NPD) measurements were performed on a 5g sample  on the diffractometer HB-2A at the High Flux Isotope Reactor (HFIR), ORNL. Measurements of the crystal structure were performed with a wavelength of $\lambda$=1.54 $\rm \AA$ and measurements of the magnetic structure were performed with $\lambda$=2.41 $\rm \AA$. Temperature measurements at select $2\Theta$ were performed with the HB-1A fixed incident energy triple-axis spectrometer with  $\lambda$=2.36 $\rm \AA$. The branching ratio of the L-edges was measured with x-ray absorption near edge spectroscopy (XANES) at beamline 4-ID-D at the Advanced Photon Source (APS), Argonne National Laboratory. The fluorescence was collected with a detector placed upstream from the sample at 22$^\circ$ from the incoming beam. Measurement were performed on different areas of the powder-on-tape sample to ensure reproducibility. The DFT calculations were performed with the generalized gradient approximation and projector augmented wave (PAW) approach \cite{BlochlDFT} as implemented in the Vienna ab initio simulation package (VASP) \cite{KresseJoubDFT,KresseFurthDFT}. For Rh and O standard potentials were used (Rh and O in the VASP distribution), and for Sr a potential in which semicore s and p states are treated as valence states is used (Srsv). The structural optimization was done using the doubled unit cell with the experimental lattice constants, a 2x2x2 k-point grid and an energy cutoff of 550 eV. Subsequently, the magnetic ground state was examined  including the local $U$ for the Rh d states to account for strong correlation effects \cite{LiechtensteinDFT} with $U$=2.5 and J$\rm _H$=0.9 eV, where J$\rm _H$ is the Hund's coupling \cite{PhysRevLett.101.026408}, as well as the SOC.

\begin{figure}[tb]
	\centering     
		\includegraphics[trim=0cm 0.5cm 0cm 2.4cm,clip=true, width=0.95\columnwidth]{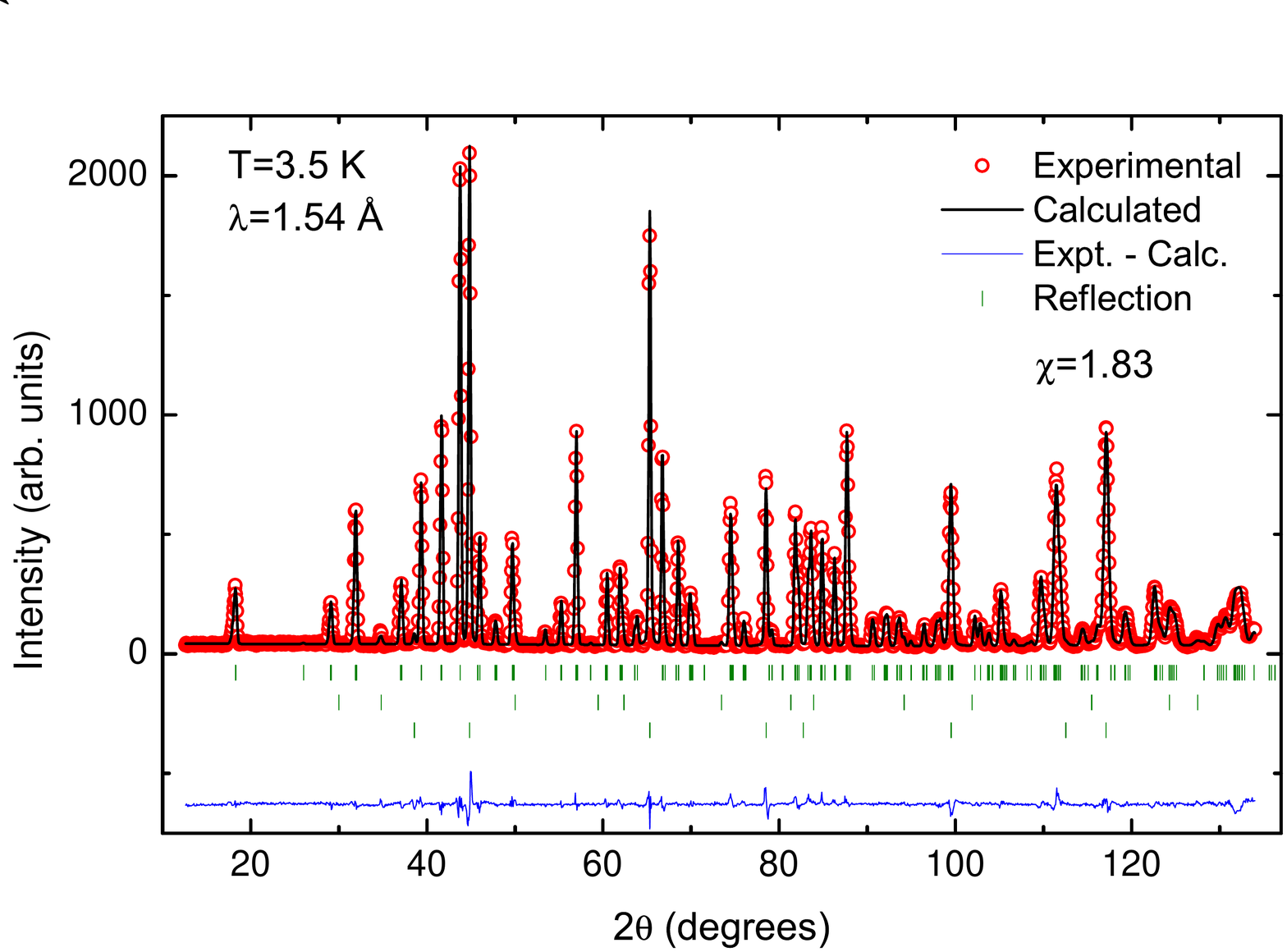}
	\includegraphics[trim=6.0cm 9.2cm 5.5cm 8.7cm,clip=true, width=0.6\columnwidth]{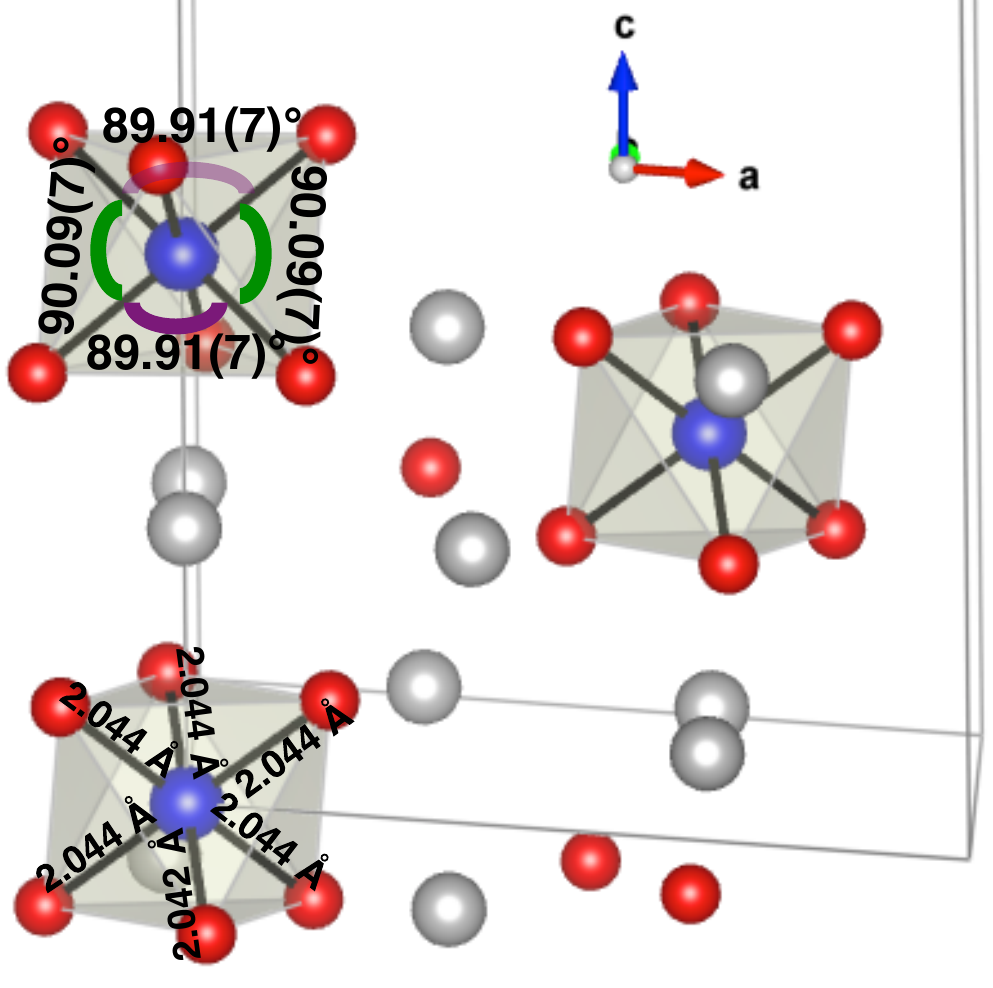}
	\caption{\label{FigStructure} Neutron powder diffraction measurements of the crystal structure of Sr$_4$RhO$_6$. The green tick marks correspond to Sr$_4$RhO$_6$, a small impurity phase of SrO and the Al sample can from top down, respectively. The bond angles and distances are shown between the O (red) and Rh (blue) ions in Sr$_4$RhO$_6$, revealing near ideal and isolated octahedra. The Sr ions are shown as grey spheres.}
\end{figure}

The structure of Sr$_4$RhO$_6$ has been previously established by laboratory x-ray measurements \cite{VenteSr416}, however this technique is appreciably less sensitive to oxygen positions compared to neutrons. Given the importance of octahedral distortions in potentially controlling the mixing and subsequent ${J}_{\mathrm{eff}}\mathbf{=}1/2$ character and emergent magnetic and electronic properties we performed NPD on the HB-2A diffractometer. The refined structure based on those in the literature remained stable to 3 K, within the magnetic regime. The sample was found to be of high quality with no Sr$_2$RhO$_4$ that forms in the intermediate growth phase and only $\sim$1.3$\%$ non-magnetic SrO impurity phase. The crystal structure obtained from NPD is shown in Fig.~\ref{FigStructure}. The disconnected RhO$_6$ octahedra are evident, with no shared Rh-O bonds between octahedra. Moreover the ideal cubic nature of these octahedra are revealed with identical Rh-O bonds (2.0437(14) $\rm \AA$) and two O-Rh-O bond angles that deviate by a remarkably small amount of less than 0.1$^\circ$ from the ideal 90$^\circ$. This contrasts with a deviation of $\sim$2$^\circ$ in Ca$_4$IrO$_6$ that is considered one of the closest  ${J}_{\mathrm{eff}}\mathbf{=}1/2$ with nearly regular octahedra \cite{PhysRevB.89.081104}. Therefore structurally Sr$_4$RhO$_6$ appears well suited to host electronic and magnetic behavior that can be controlled by SOC.

\begin{figure}[tb]
	\centering     
	\includegraphics[trim=2.5cm  0.5cm 2.2cm 1.0cm,clip=true, width=0.7\columnwidth]{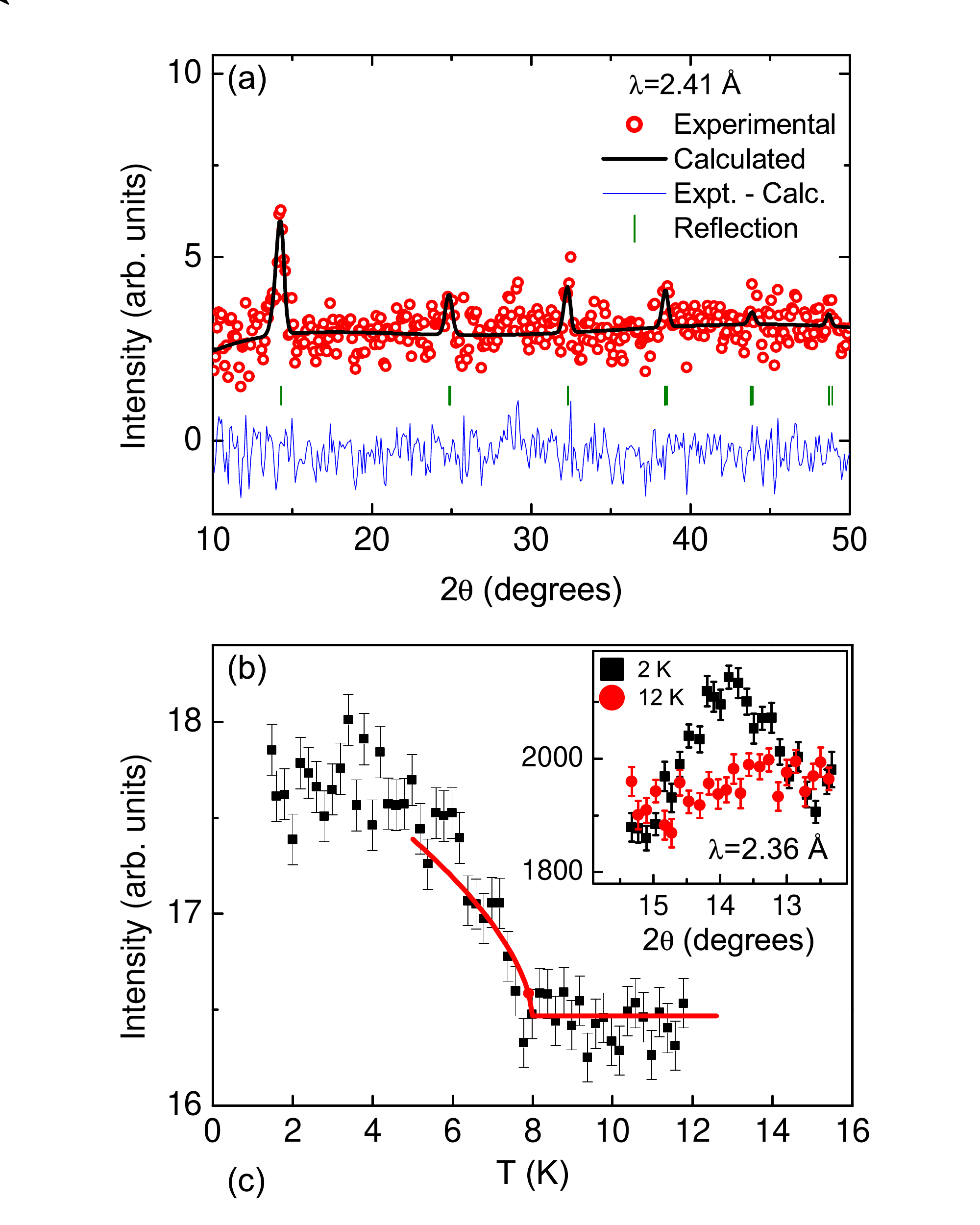} 
		\includegraphics[trim=5.5cm  9.0cm 7.7cm 12.5cm,clip=true, width=0.7\columnwidth]{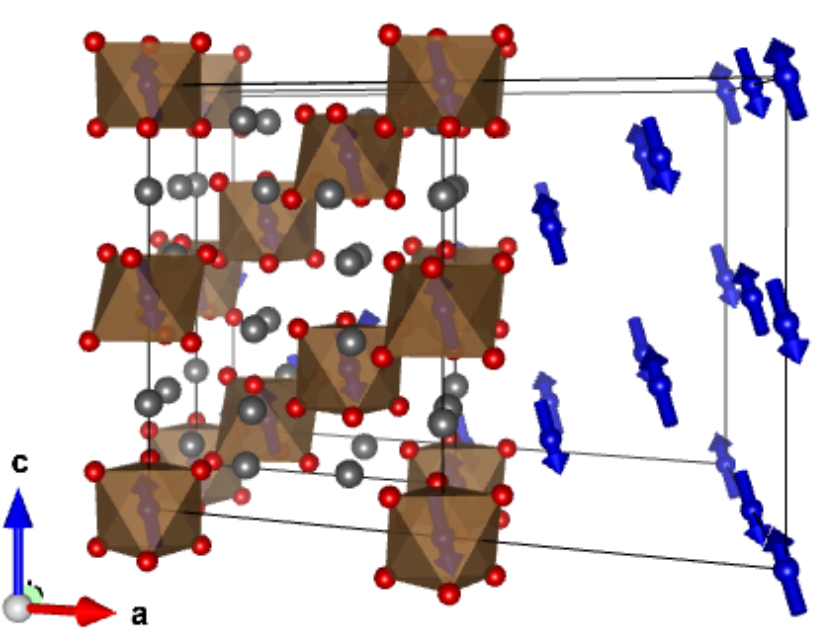}
	\caption{\label{FigMagNeutron} (a) Difference of 15 K and 3 K neutron powder diffraction measurements. The fit is to the $\Gamma_1$ IR magnetic model. (b) The intensity of a magnetic reflection reveals long range magnetic ordering at 7.4(5) K in Sr$_4$RhO$_6$, the fit is to a power law. (c) The lowest energy $\Gamma_1$ magnetic ordering in Sr$_4$RhO$_6$.}
\end{figure}

Before considering the electronic behavior further we address the significant question of whether Sr$_4$RhO$_6$ orders magnetically and the nature of the ground state. NPD has proven challenging for ${J}_{\mathrm{eff}}\mathbf{=}1/2$ Ir-based compounds due to the combination of large neutron absorption, small moment size and pronounced magnetic form factor induced reduction in intensity with scattering angle. All of these factors are reduced when moving from Ir to Rh based systems and this allowed a successful NPD investigation with measurements performed through the reported anomaly in susceptibility to probe the microscopic magnetic structure. Comparing results at 15 K and 3 K shows the presence of extra intensity at low temperature at several reflection positions, see Fig.~\ref{FigMagNeutron}(a). All of the reflections are consistent with a propagation vector of ${\bf k}$$=$($\frac{1}{2} \frac{1}{2} 0$) which corresponds to the same ordering vector as found in the iridate analogue Ca$_4$IrO$_6$ \cite{PhysRevB.89.081104}. Following a representational analysis approach yields two possible irreducible representations (IR) that are both consistent with the observed NPD measurements \cite{sarahwills}. These correspond to $\Gamma_1$ and $\Gamma_3$ in Kareps numbering scheme for the Rh ion at the $6b$ Wyckoff position, (0,0,0) site. Both IRs describe similar antiferromagnetic structures, with the distinction being either antiferromagnetic spins along the c-axis for $\Gamma_1$ or ferromagnetic chains along the c-axis for $\Gamma_3$. Given the non-first order nature of the transition (see Fig.~\ref{FigMagNeutron}(b)) we employ the simplification that only one IR describes the magnetic structure. To uncover the lowest energy magnetic ground state we performed DFT calculations for Sr$_4$RhO$_6$.  It is found that the $\Gamma_1$  structure is lower in energy than the $\Gamma_3$  structure by 0.24eV per magnetic unit cell (0.01eV per Rh). Starting from parameters based on the refined NPD measurements we find $\Gamma_1$ with spins predominantly along the c-axis to be the lowest energy ground state. The results of magnetic refinements of the NPD for $\Gamma_1$ are shown in Fig.~\ref{FigMagNeutron}(a). The ordered magnetic moment from NPD measurements is 0.66(5)$\rm \mu_B$/Rh ion. Following the intensity of one of the magnetic reflections with temperature yields a magnetic ordering temperature of 7.4(5) K, as shown in Fig.~\ref{FigMagNeutron}(b). Contrasting with results on Ca$_4$IrO$_6$ \cite{PhysRevB.89.081104} reveals both compounds adopt the same magnetic structure, although with slightly different canting from the c-axis and although the moment sizes are similar it is slightly larger in the rhodate. This would be expected in two compounds with similar magnetic ordering temperatures and underlying behavior, with the principle distinction being reduced itinerancy in the rhodate and subsequent increased local moment.

\begin{figure}[tb]
	\centering     
	\includegraphics[trim=0.2cm  0.0cm 0cm 0.0cm,clip=true, width=0.8\columnwidth]{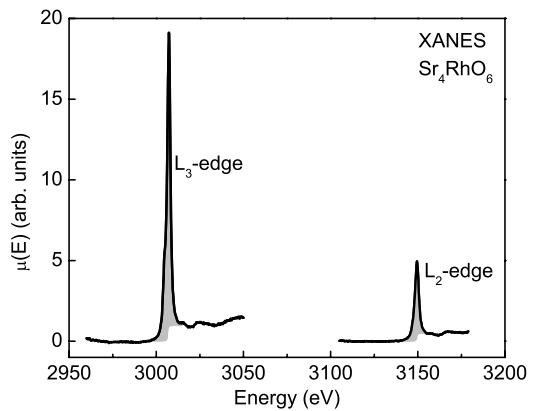}
	\caption{\label{FigXANES} Normalized and self-absorption corrected XANES fluorescence measurements through the L$_2$ and L$_3$ edges in Sr$_4$RhO$_6$. The peak in the white line occurs at L$_2$=3149.0 eV and L$_3$=3007.12 eV. Measurements were performed at room temperature.}
\end{figure}

With the similarities between Sr$_4$RhO$_6$ and the ${J}_{\mathrm{eff}}\mathbf{=}1/2$ iridate Ca$_4$IrO$_6$ in terms of the magnetic structure we now experimentally  consider the role of SOC on Sr$_4$RhO$_6$. A useful probe in this regard for the iridates has proven to be x-ray absorption spectroscopy, where the branching ratio (BR) of the L-edges provides evidence for the existence of enhanced SOC if the ratio deviates from the statistical value of 2 \cite{PhysRevLett.105.216407}. The BR is obtained by fitting the white lines from XANES measurements to a resolution broadened step function plus a lorentzian to obtain the integrated intensity, shown as the grey area in Fig.~\ref{FigXANES}. The results reveal BR=3.6(2) for Sr$_4$RhO$_6$. The strong deviation from statistical indicates an appreciable role for SOC in the ground state. Indeed the values correspond closely to the BR for iridates that show strongly SOC enhanced behavior  \cite{PhysRevLett.105.216407}. Similar XANES measurements on $\alpha$-RuCl$_3$ revealed a BR of 3 that was presented as evidence for substantial SOC \cite{PhysRevB.90.041112}. 

\begin{figure}[tb]
	\centering     
	\includegraphics[trim=0.44cm 4.3cm 1.0cm 2.4cm,clip=true, width=1.0\columnwidth]{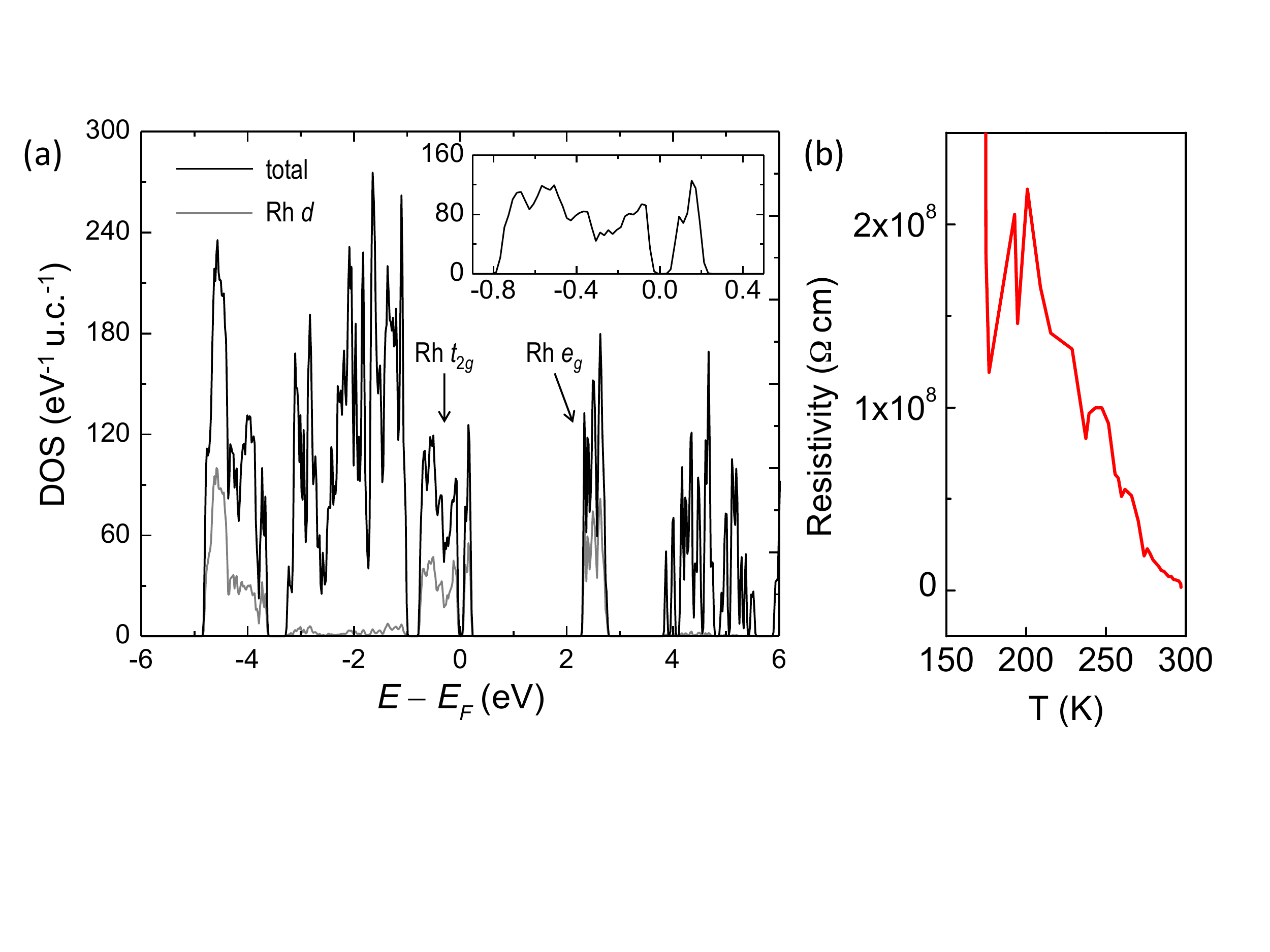} 
	\caption{\label{FigDFT} (a) DFT calculations show the $\Gamma_1$ magnetic ordering is the lowest energy and insulating in Sr$_4$RhO$_6$.  The area region around the Fermi energy is shown inset (b) The insulating nature was experimentally probed with resistivity measurements. The resistance increased with decreasing temperature, becoming immeasurably high below 160 K.}
\end{figure}

To further investigate the role of SOC we performed DFT calculations on Sr$_4$RhO$_6$ including the $\Gamma_1$ magnetic ground state. The results (Fig.~\ref{FigDFT}(a)) show a $t_{2g}$ manifold split into  ${J}_{\mathrm{eff}}\mathbf{=}1/2$ and ${J}_{\mathrm{eff}}\mathbf{=}3/2$ dominated bands. The  ${J}_{\mathrm{eff}}\mathbf{=}1/2$ states are higher in energy then ${J}_{\mathrm{eff}}\mathbf{=}3/2$ with the near-Fermi-level states given predominantly by the ${J}_{\mathrm{eff}}\mathbf{=}1/2$ states, and ${J}_{\mathrm{eff}}\mathbf{=}3/2$ fully filled. Therefore, while the splitting between ${J}_{\mathrm{eff}}\mathbf{=}1/2$ and ${J}_{\mathrm{eff}}\mathbf{=}3/2$ is not complete, as can be seen in the DOS, spectroscopically a pure ${J}_{\mathrm{eff}}\mathbf{=}1/2$ is realized because the scattering process is an excitation to the unoccupied states. Experimentally we probed the insulating nature with resistance measurements on a pressed pellet, see Fig.~\ref{FigDFT}(b). These results are consistent with an insulating material and moreover the powder is dull, characteristic of an insulator. From DFT calculations an insulating gap of around 0.1 eV is observed. This insulating gap is reduced from that found in Ca$_4$IrO$_6$ from DFT of 0.6 eV. Moreover we find that unlike Ca$_4$IrO$_6$ the ${J}_{\mathrm{eff}}\mathbf{=}1/2$ and ${J}_{\mathrm{eff}}\mathbf{=}3/2$ bands are not fully separated. This mirrors the situation from DFT calculations in the paramagnetic fluoride based Ir and Rh systems predicted to both be ${J}_{\mathrm{eff}}\mathbf{=}1/2$ Mott insulators \cite{PhysRevLett.114.096403}. The mixed ${J}_{\mathrm{eff}}$ bands indicates that despite the essential ideal octahedra in Sr$_4$RhO$_6$ the intrinsic SOC is not large enough to fully split the $t_{2g}$ manifold. This situation is analogous to several other candidate ${J}_{\mathrm{eff}}\mathbf{=}1/2$ materials where the opposite occurs: the ${J}_{\mathrm{eff}}\mathbf{=}1/2$ and ${J}_{\mathrm{eff}}\mathbf{=}3/2$ bands are mixed despite the large SOC due to appreciable non-cubic distortions. 

Collectively our experimental and theoretical results reveal the behavior in Sr$_4$RhO$_6$ is strongly influenced by SOC. Qualitatively the results are very similar to the Rh based fluorides, however it is important to consider the differing effects of oxygen compared to fluorine. Specifically in oxides the hybridization is generally much stronger than fluorides. Consequently the band width of ${J}_{\mathrm{eff}}\mathbf{=}1/2$ dominated bands and ${J}_{\mathrm{eff}}\mathbf{=}3/2$ dominated bands are narrower in fluorides, of the order 0.1 eV compared to 0.5 eV in Sr$_4$RhO$_6$. Therefore while the degree of separation is stronger in the fluorides both show ${J}_{\mathrm{eff}}\mathbf{=}1/2$ character, although neither show fully separated ${J}_{\mathrm{eff}}\mathbf{=}1/2$ and ${J}_{\mathrm{eff}}\mathbf{=}3/2$ dominated bands. This leads to the question of whether an alternative approach is more appropriate in which both these fluoride and oxides system are best described by three Kramers states as considered for other rhodates \cite{doi:10.1021/ic402653f}. However, as suggested in Ref.~\onlinecite{doi:10.1021/ic402653f} the BR from XANES should distinguish between these cases and indeed this does so in Sr$_4$RhO$_6$.

In conclusion, we have observed a rare occurrence of long range magnetic order in a rhodate compound. Moreover, Sr$_4$RhO$_6$ is found to be an insulator, leading to a consideration in terms of SOC enhanced behavior. Experimentally, the magnetic ordering and insulating behavior are all analogous to the isostuctural ${J}_{\mathrm{eff}}\mathbf{=}1/2$ iridate Ca$_4$IrO$_6$. DFT calculations reveal
Jeff character with mixed ${J}_{\mathrm{eff}}\mathbf{=}1/2$ and ${J}_{\mathrm{eff}}\mathbf{=}3/2$ dominated
bands. Despite the mixing due to the reduced SOC in going
from Ir to Rh, the physical properties are strongly influenced by
SOC. Therefore, with Sr$_4$RhO$_6$ being shown to be a 4d TMO
with ${J}_{\mathrm{eff}}$ character, further investigations on oxides with similar
spatially disconnected octahedra offer routes to uncovering
analogous exotic properties as found in the iridates.


\begin{acknowledgments}
This research at ORNL's High Flux Isotope Reactor was sponsored by the Scientific User Facilities Division, Office of Basic Energy Sciences, U.S. Department of Energy. Use of the Advanced Photon Source, an Office of Science User Facility operated for the U.S. DOE Office of Science by Argonne National Laboratory, was supported by the U.S. DOE under Contract No. DE-AC02-06CH11357. S.O. is supported by the U.S. Department of Energy, Office of Science, Basic Energy Sciences,Materials Sciences and Engineering Division.
\end{acknowledgments}


%

\end{document}